\documentclass[12pt]{article}
\usepackage{graphicx}

\usepackage{xspace}
\usepackage[dvipsnames]{xcolor}
\definecolor{bleuCMS}{RGB}{0,124,174}
\definecolor{bleuCMSBg}{RGB}{0,165,232}
\definecolor{bleuCMS!40!bleuCMSBg}{RGB}{0,124,174}

\usepackage[sorting=none]{biblatex} 
\newcommand{\pt}{\ensuremath{p_{\mathrm{T}}}\xspace}

\newcommand{\fbinv}{\ensuremath{\mathrm{\,fb^{-1}}}\xspace}
\newcommand{\ttbar}{\ensuremath{\mathrm{t\overline{t}}}\xspace}
\newcommand{\JPsi}{\ensuremath{J/\psi}\xspace}
\newcommand{\Dz}{\ensuremath{\mathrm{D^{0}}}\xspace}
\newcommand{\Dzm}{\ensuremath{\mathrm{D^{0}_{\mu}}}\xspace}
\newcommand{\rb}{\ensuremath{r_{\mathrm{b}}}\xspace}
\newcommand{\stat}{\ensuremath{\mathrm{\,(stat)}}\xspace}
\newcommand{\syst}{\ensuremath{\mathrm{\,\xspace(syst)}}\xspace}
\def\sPlot{\hbox{$_s$}{\cal P}lot}

\newcommand{\fit}{\mbox{\ensuremath{\rb=0.855 \pm 0.037 \stat \pm 0.031 \syst}}}
\newcommand{\jpsistat}{\mbox{\ensuremath{\rb=0.864 \pm 0.053 \stat}}}
\newcommand{\dzstat}{\mbox{\ensuremath{\rb=0.836 \pm 0.070\stat}}}
\newcommand{\dzmustat}{\mbox{\ensuremath{\rb=0.858 \pm 0.072 \stat}}}
\newcommand{\jpsifit}{\mbox{\ensuremath{\jpsistat \pm 0.040 \syst}}}
\newcommand{\dzfit}{\mbox{\ensuremath{\dzstat \pm 0.056 \syst}}}
\newcommand{\dzmufit}{\mbox{\ensuremath{\dzmustat \pm 0.081 \syst}}}

\newcommand\pubnumber{Experiment-1-1}
\newcommand\pubdate{\today}

\def\institute{Department of Physics\\
The Ohio State University}

\def\Title#1{\begin{center} {\Large #1 } \end{center}}
\def\Author#1{\begin{center}{ \sc #1} \end{center}}
\def\Address#1{\begin{center}{ \it #1} \end{center}}

\newcommand\pubblock{\rightline{\begin{tabular}{l} \pubnumber\\
         \pubdate  \end{tabular}}}
\newenvironment{Abstract}{\begin{quotation}  }{\end{quotation}}
\newenvironment{Presented}{\begin{quotation} \begin{center} 
             PRESENTED AT\end{center}\bigskip 
      \begin{center}\begin{large}}{\end{large}\end{center} \end{quotation}}

\def\beq{\begin{equation}}
\def\eeq#1{\label{#1}\end{equation}}
\def\eeqn{\end{equation}}

\def\beqa{\begin{eqnarray}}
\def\eeqa#1{\label{#1}\end{eqnarray}}
\def\eeqan{\end{eqnarray}}

\let\bar=\overbar

\def\Dslash{\not{\hbox{\kern-4pt $D$}}}
\def\dslash{\not{\hbox{\kern-2pt $\del$}}}

\def\msb{{\bar{\ssstyle M \kern -1pt S}}}

\begin{document}
\begin{titlepage}
\pubblock

\vfill
\Title{YSF Measurement of the shape of the $\mathrm{b}$ quark fragmentation function using charmed mesons produced inside $\mathrm{b}$ jets from $\ttbar$ decays}
\vfill
\Author{Brent R. Yates}
\Address{\institute}
\vfill
\begin{Abstract}
A first measurement of the $\mathrm{b}$ quark fragmentation function at the LHC is presented. Charmed meson candidates produced via semi-leptonic $\ttbar$ decays ($\mathrm{t} \to \mathrm{b}\mathrm{W} \to$ $\mathrm{b}$-jet $\ell \nu$) are used as a proxy for the parent $\mathrm{B}$ meson. Templates are generated at various values of the fragmentation shape parameter $\rb$, and are fit to the data to measure the value of $\rb$. The final fit result is $\fit$. This is the first measurement of the $\mathrm{b}$ quark fragmentation function within a color rich environment, and is consistent with previous results from $e^+e^-$ colliders.
\end{Abstract}
\vfill
\begin{Presented}
$14^\mathrm{th}$ International Workshop on Top Quark Physics\\
(videoconference), 13--17 September, 2021
\end{Presented}
\vfill
\end{titlepage}
\def\thefootnote{\fnsymbol{footnote}}
\setcounter{footnote}{0}

\section{Introduction}
The shape of the fragmentation for b~quark decays is a fundamental piece of modeling many physics processes at the Large Hadron Collider (LHC)~\cite{Bruning:2004ej,Evans:2008zzb}. This means the shape must be measured as precisely as possible. Because the hadronization of quarks is very complicated, no exact functional form exists. Instead, we must rely on empirical models to approximate the behavior. The simulation software PYTHIA~\cite{Sjostrand:2007gs} provides a few different models. The default used on the Compact Muon Solenoid (CMS) Experiment~\cite{CMS:2008xjf} is the Lund-Bowler model. This model is an extension to the Lund symmetric model, which accounts for the finite mass of the bottom (b) and and charm mesons (c) to account for the slightly softer distributions observed in experimental data. To date, the best measurement of the b~quark fragmentation function~\cite{Skands:2014pea} comes from a combination of $e^+e^-$  colliders, including the Large Electron-Positron Collider (LEP)~\cite{DELPHI:2011aa,Heister:2001jg,Abbiendi:2002vt} formerly at CERN, and the SLAC Linear Collider (SLC)~\cite{Abe:2002iq} former at SLAC. The LHC, being a proton-proton collider, produces a color-right environment, differing from the $e^+e^-$ colliders. Therefore, the natural question is, do we expect the fragmentation function to be fundamentally different in this type of environment? This proceeding will summarize the latest result form the CMS Collaboration, providing the first direct measurement of the b~quark fragmentation function within \ttbar events.

\section{The analysis}
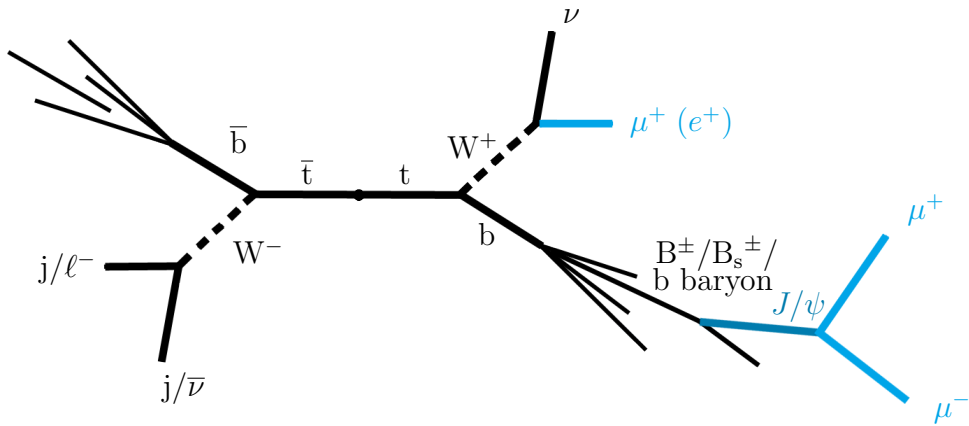
\begin{figure}[!htb]
\centering
\includegraphics[width=12cm]{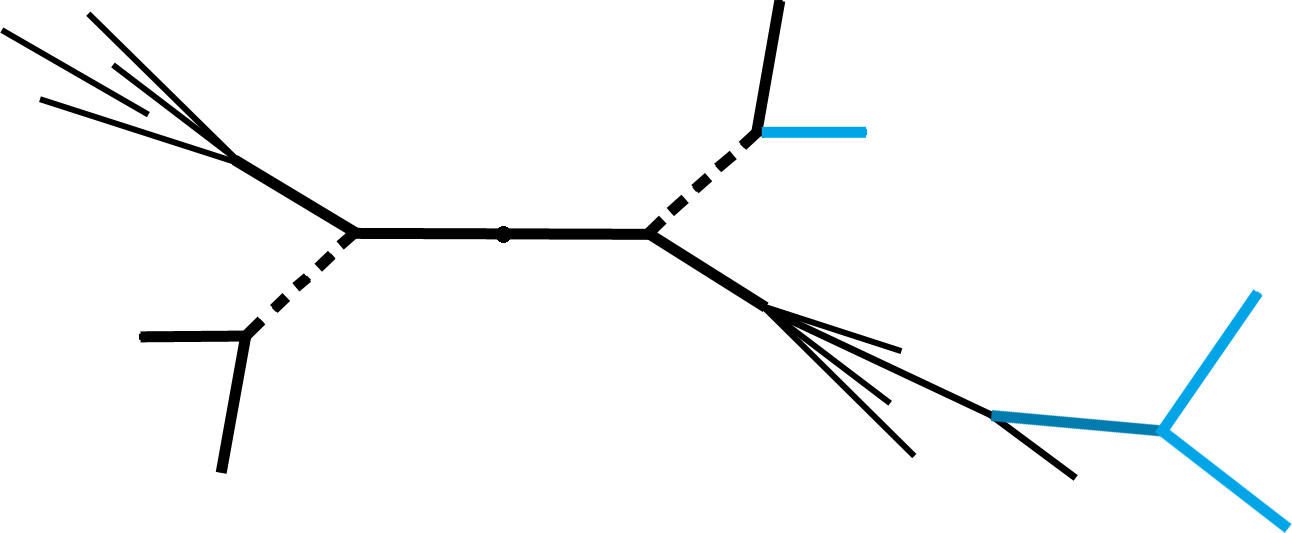}
\put(-269, 83){\makebox[0.1\textwidth][r]{$\mathrm{\overline{t}}$}}
\put(-294, 95) {\makebox[0.1\textwidth][r]{$\mathrm{\overline{b}}$}}
\put(-280, 53) {\makebox[0.1\textwidth][r]{$\mathrm{W^{-}}$}}
\put(-351, 48) {\makebox[0.1\textwidth][r]{$\mathrm{j}/\ell^{-}$}}
\put(-310, 3){\makebox[0.1\textwidth][r]{$\mathrm{j}/\overline{\nu}$}}
\put(-232, 83) {\makebox[0.1\textwidth][r]{$\mathrm{t}$}}
\put(-199, 93) {\makebox[0.1\textwidth][r]{$\mathrm{W^{+}}$}}
\put(-200, 60){\makebox[0.1\textwidth][r]{$\mathrm{b}$}}
\put(-93, 53){\makebox[0.1\textwidth][r]{$\mathrm{B^{\pm}}/\mathrm{B_{s}}^{\pm}/$}}
\put(-95, 43){\makebox[0.1\textwidth][r]{$\mathrm{b}$ baryon}}
\put(-168, 144) {\makebox[0.1\textwidth][r]{$\nu$}}
\put(-110, 103) {\makebox[0.1\textwidth][r]{\textcolor{bleuCMSBg}{$\mu^{+}\;(e^{+})$}}}
\put(-75, 34) {\makebox[0.1\textwidth][r]{\textcolor{bleuCMS!40!bleuCMSBg}{\JPsi}}}
\put(-30, 70) {\makebox[0.1\textwidth][r]{\textcolor{bleuCMSBg}{$\mu^{+}$}}}
\put(-20, -5) {\makebox[0.1\textwidth][r]{\textcolor{bleuCMSBg}{$\mu^{-}$}}}
\caption{Pictorial view of an exclusive $\JPsi$ production in a $\ttbar$ system.}
\label{fig:channel}
\end{figure}

The main focus of this analysis is to measure the b~quark fragmentation function using charm mesons as a proxy for the parent B~hadrons. This proxy is defined as
\begin{equation}
    \frac{\mathrm{charm\;meson}\;\pt}{\sum \pt^{ch}},
\end{equation}
where \pt is the transverse momentum of the charm meson, and $\sum \pt^{ch}$ is the scalar sum of the transverse momentum of all the charged particles within the b~jet containing the charm meson. This analysis is restricted to charged particles in order to bypass many of the experimental uncertainties associated with measuring neutral particles in the CMS detector. A Kalman Vertex Fit (KVF)~\cite{Speer:1364622} is used to reliably identify b~jets produced in \ttbar decays. This is because the top quark will decay almost 100\% of the time to a $\mathrm{b}$ quark via the weak nuclear force, resulting in a W boson decaying into either two quarks---which are resolved in the CMS detector as jets---or into a charged lepton ($\ell$) and its associated anti-neutrino ($\overline{\nu}_{\ell}$), as shown in Fig.~\ref{fig:channel}. This analysis focuses on the semi-leptonic and di-leptonic decay channels, using $e$ and $\mu$ as the leptons. The specific charm mesons considered are the $\JPsi$ and $\Dz$. The $\Dz$ channel is sub-divided into events with an additional soft $\mu$, labeled $\Dzm$, and events without the additional soft $\mu$. The full 2016 dataset is used, with an integrated luminosity of 36\fbinv.

\section{Signal extraction}
\begin{figure}
    \centering
    \includegraphics[width=0.3\textwidth]{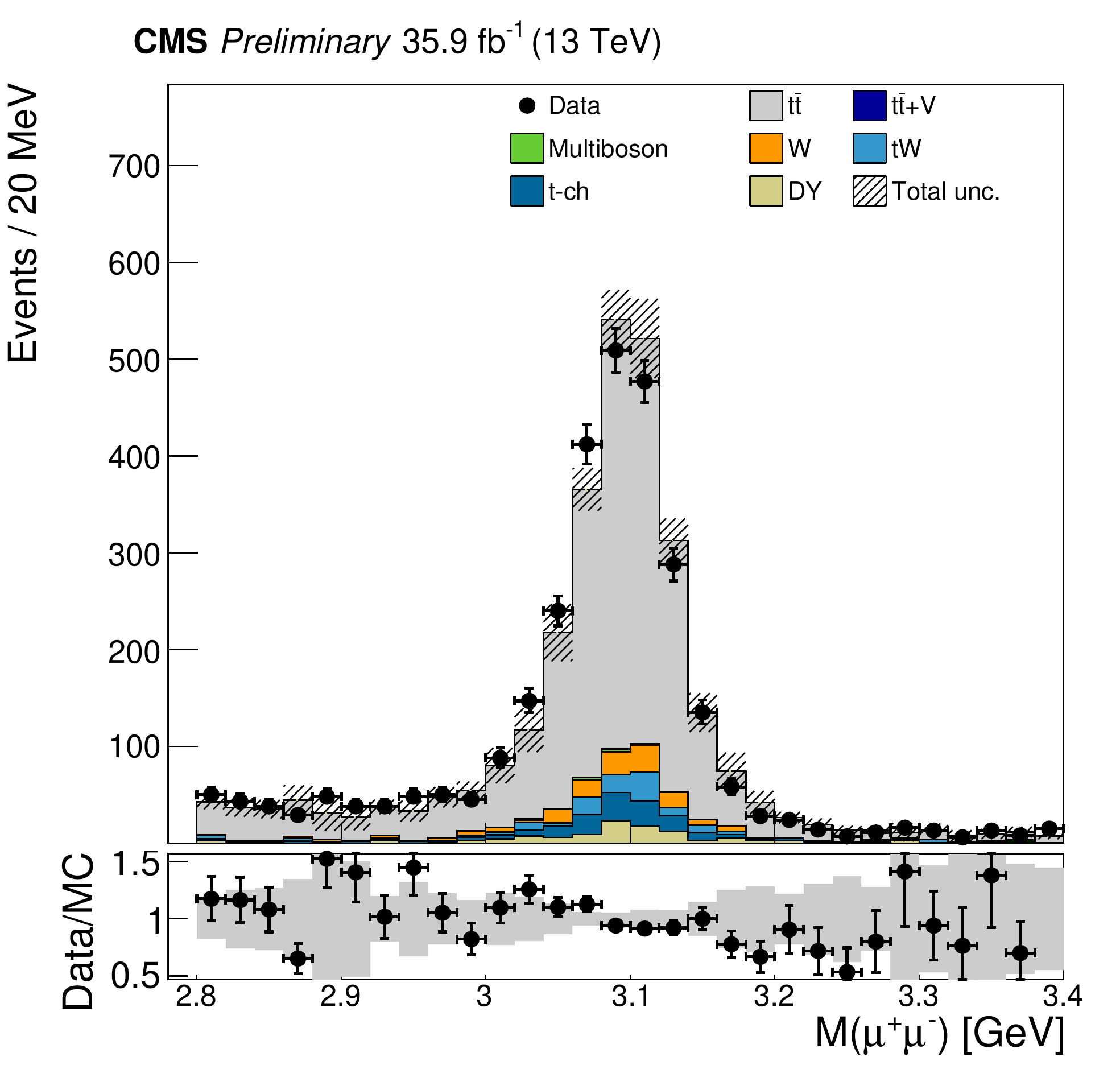}
    \includegraphics[width=0.3\textwidth]{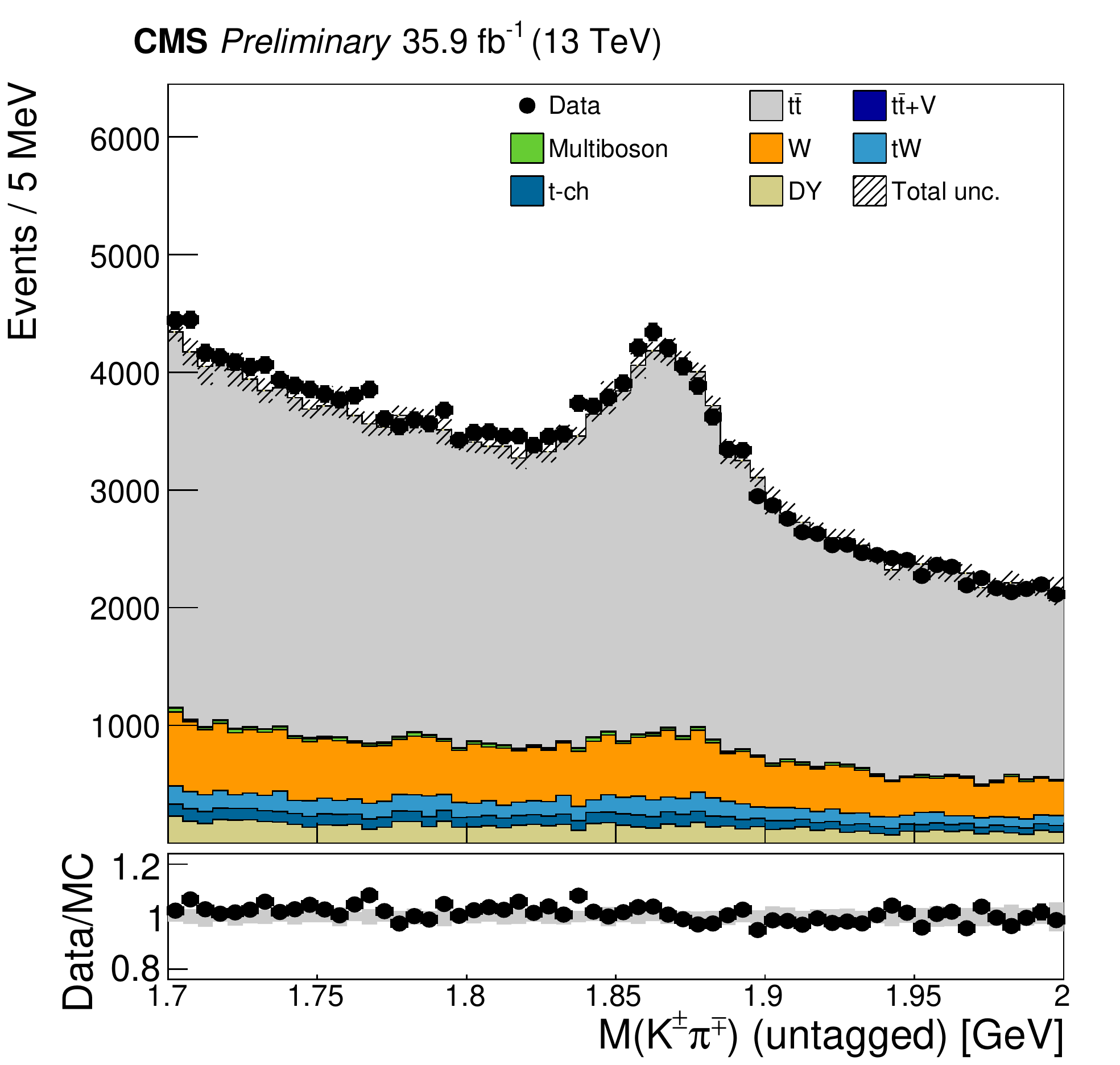}
    \includegraphics[width=0.3\textwidth]{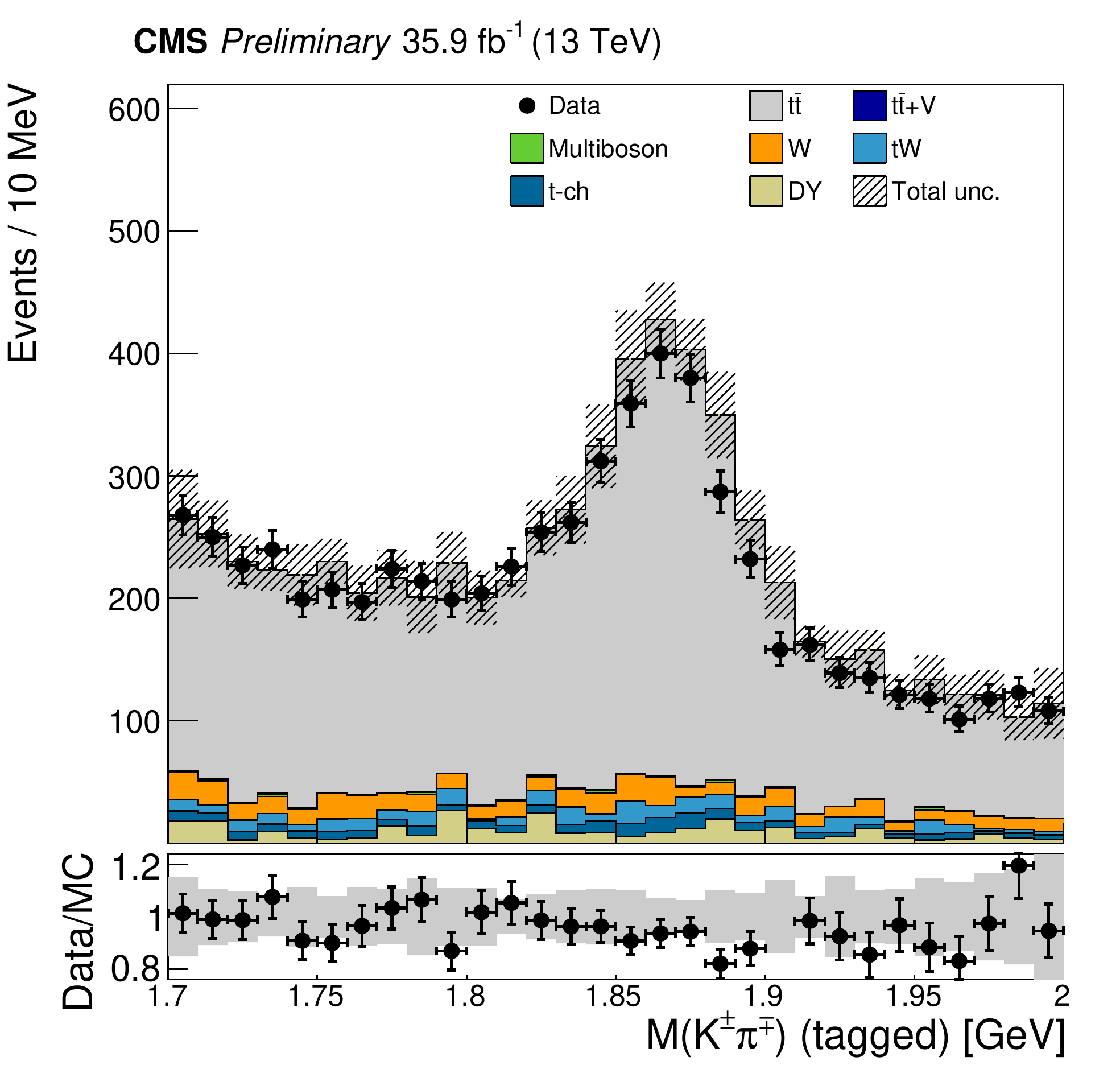}
    \caption{Invariant mass plots for the \JPsi, \Dz, and \Dzm mesons.}
    \label{fig:inv_mass}
\end{figure}

A maximum likelihood fit is performed on the invariant mass of the three charm mesons as shown in Fig.~\ref{fig:inv_mass}. The fit for the \JPsi consists of a Crystal Ball function for the signal region, and an exponential function for the small combinatorial background. The \Dz and \Dzm masses are fit using a Gaussian for each signal, and an exponential function for the background. The two \Dz mesons also contain an additional background Gaussian function to capture the Cabibbo suppressed $\Dz \to \mathrm{K^{+}K^{-}}$ in which one K is misreconstucted as a $\pi$. These three invariant mass fits are then passed to $\sPlot$~\cite{Pivk:2004ty} in order to extract a probability for each event to be signal or background. Finally, additional background subtractions are performed directly on the charm meson proxy distributions to remove any contamination from c~jet, Drell-Yan, and single~t quark production.

\section{Fitting procedure}
\begin{figure}[!htb]
    \centering
    \includegraphics[width=0.45\textwidth]{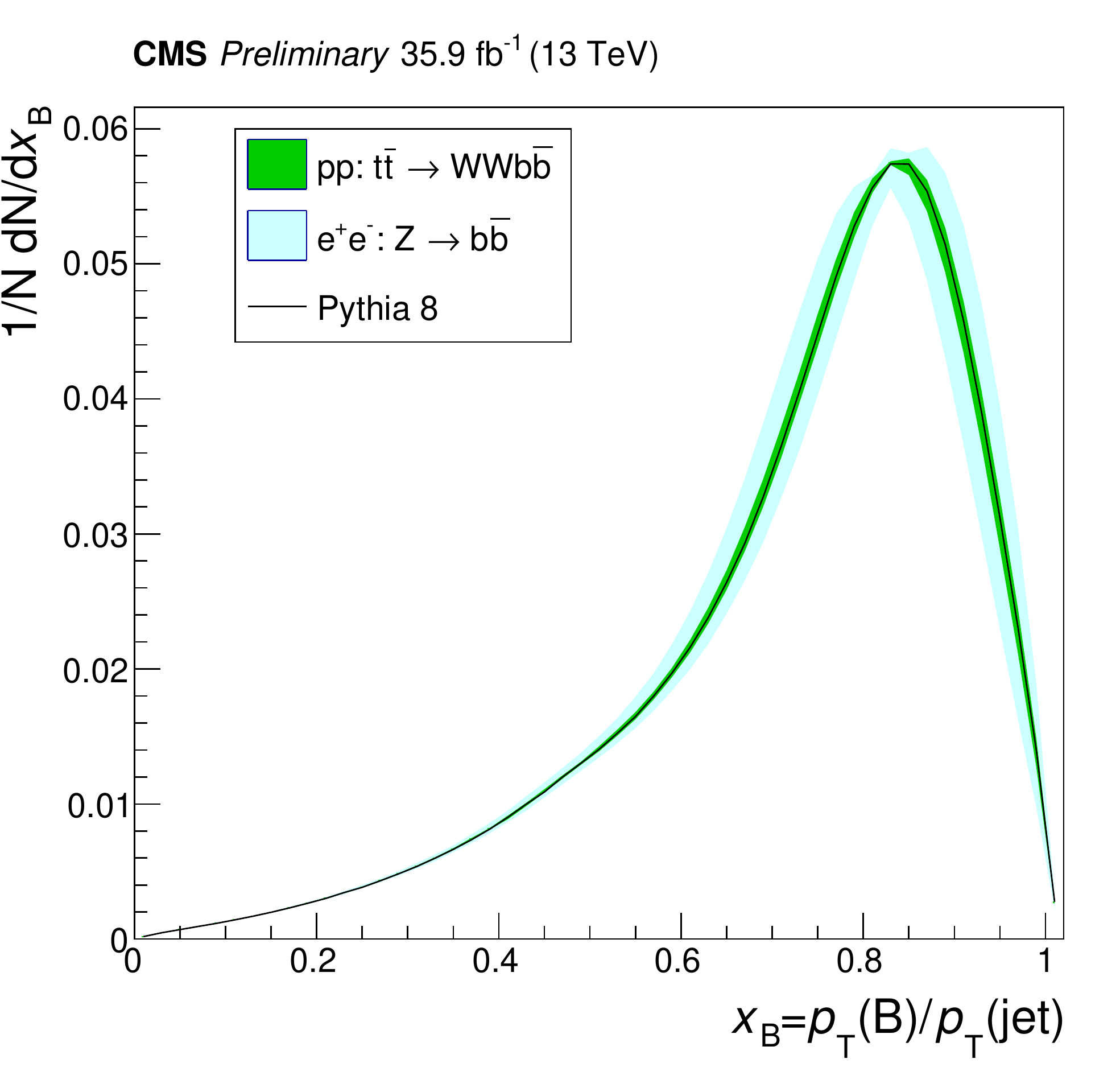}
    \includegraphics[width=0.45\textwidth]{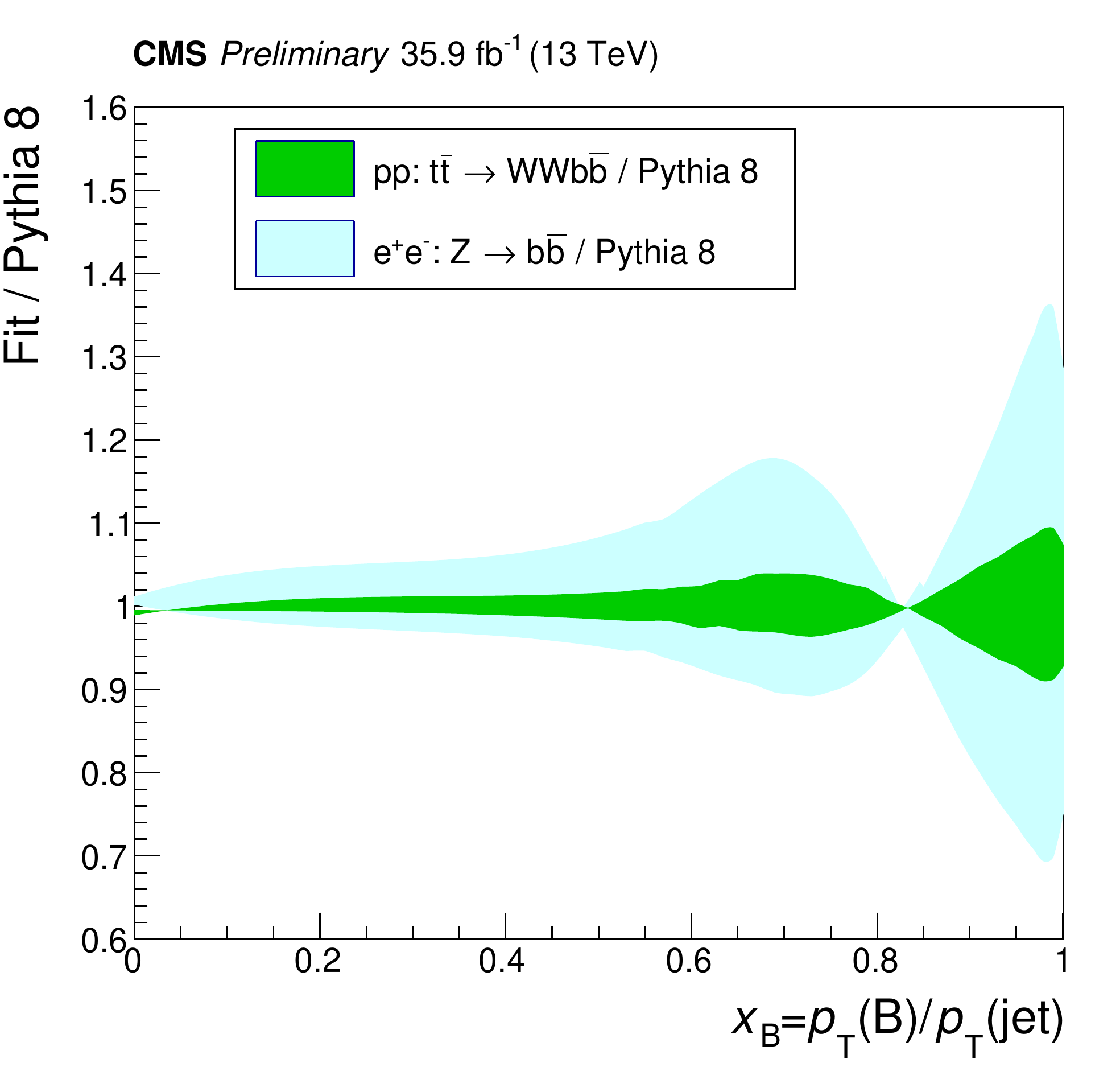}
    \caption{A comparison of the fragmentation functions (left) and the ratio of each to the nominal PYTHIA value (right).}
    \label{fig:gen_frag}
\end{figure}

Templates are generated for each proxy by varying the value of \rb in the Monte Carlo simulations. A ratio of each of these modified distributions as taken with respect to the nominal value ($\rb=0.855$), and these ratios are used as event weights to morph the simulations at the detector level. Each morphed template corresponding to an \rb value is fit to the data using a $\chi^2$ goodness-of-fit. The goodness-of-fit values are plotted, and then fit with a third order polynomial, to produce a continuous function of \rb. Finally, the minimum value of the \rb curve is extracted as the best-fit value, and the difference in \rb at $\Delta \chi^2 \pm 1$ is taken as the statistical uncertainty. Each charm meson distribution is fit independently, as well as in a simultaneous fit. The simultaneous fit is cross-checked using the BLUE method~\cite{Lyons:1988rp}. All relevant systematic uncertainties are taken into account. This is accomplished by shifting any experimental or theoretical inputs by their one standard deviation uncertainty, and evaluating the resulting shift in \rb. The final fit values are $\jpsifit$, $\dzfit$, and $\dzmufit$ for the \JPsi, \Dz, and \Dzm mesons respectively. The simultaneous fit value is\\ \fit. This is roughly a factor of four better precision than the previous $e^+e^-$ collider fit of $rb=0.894^{+0.184}_{-0.197}$. Figure~\ref{fig:gen_frag} shows a comparison between the nominal PYTHIA value, the $e^+e^-$ fit value, and the fit value of this analysis.

\printbibliography

\end{document}